# X-ray Writing of Fiber Bragg Gratings (FBGs)

Shaolin Liao[1,2] (sliao5@iit.edu)

*1Illinois Institute of Technology, 10W 31 St, Chicago, IL USA 60616*

*2Nuclear Science and Engineering, Argonne National Laboratory, Lemont, IL USA 60439*

## Abstract

Fabrication of nanoscale optical Fiber Bragg Gratings (FBGs) is one of the key manufacturing processes for fiber optics, which has many important applications in data communication and distributed remote sensing at a distance up to hundreds of kms. However, the fabrication of the FBGs is challenging. To meet such need, we propose a novel method to write the FBGs using high-flux synchrotron x-ray with a nanoscale gold/Si gratings to modulate the x-ray flux and thus change the refractive index contrast of the optical fiber to form the FBGs. The nanoscale gold/Si with high aspect ratio has been fabricated at the Center for Nanoscale Materials (CNM) at Argonne National Laboratory (ANL) and the preliminary experiment of x-ray writing of the optical fiber FBGs has been carried out at the Advanced Photon Source (APS) of ANL. The preliminary experiment shows promising result of the x-ray writing method of the optical fiber FBGs.

## 1. Theoretical Background

Fiber optics has many important applications such as high-throughput data communication and distributed remote sensing of temperature, pressure and acoustics etc. at a distance up to hundreds of kms [1-9]. However, fabrication of nanoscale Fiber Bragg Gratings (FBGs), one of the key component of fiber optics, is challenging. This is mainly due to the difficulty to change the refractive index of the optical fiber core materials that is made of silica and its dopants, especially for the active fiber doped with active elements such as Ytterbium [2, 4, 6, 8]. Here, we propose a novel method of using the high-flux synchrotron x-ray to change the refractive index of the optical fiber to form the FBGs.

The work is partly supported by National Nuclear Security Administration (NNSA) of the U.S. Department of Energy (DOE), under Contract No. DOE-NNSA PRJ1000773, Use of the Center for Nanoscale Materials, and Advanced Photon Source, Office of Science user facilities, was supported by the U.S. DOE, Office of Science, Office of Basic Energy Sciences, under Contract No. DE-AC02-06CH11357.

It is well known that the pitch/period of the FBGs $\Lambda$ is decided by the intrinsic vacuum wavelength of interest $\lambda_0$ as follows,

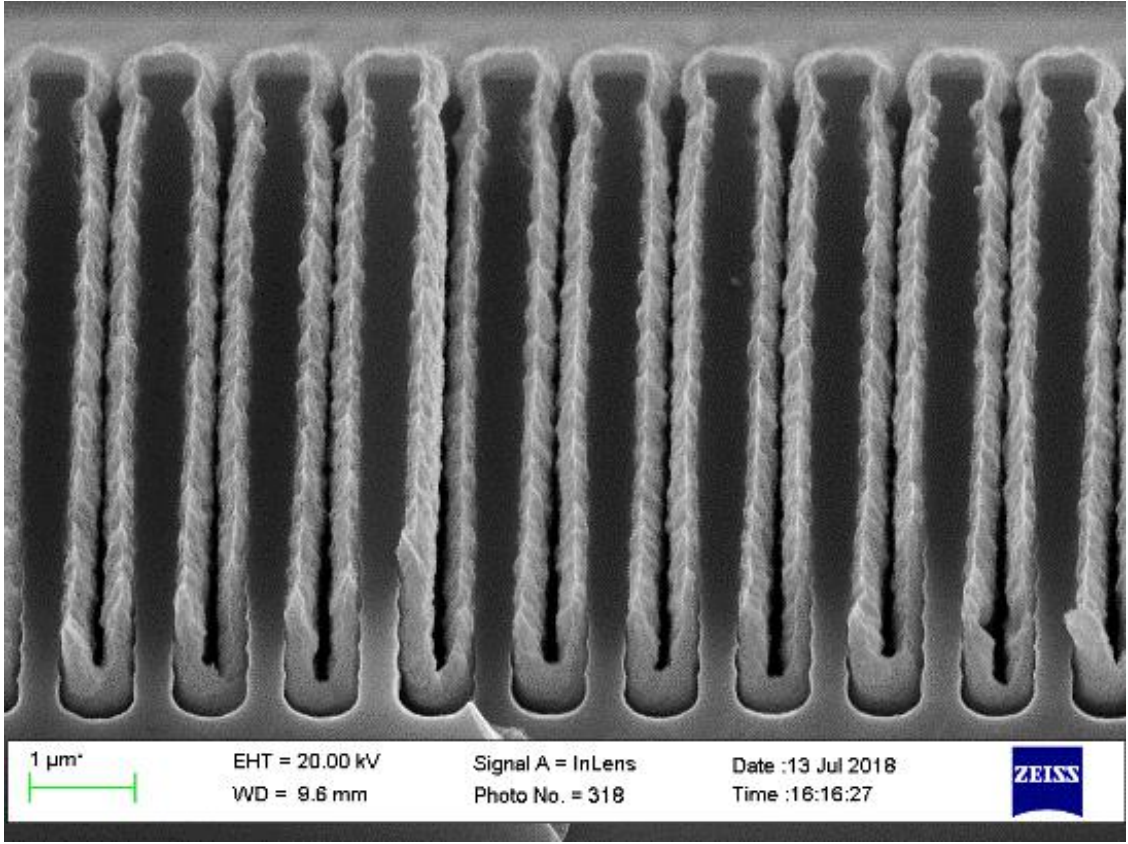

Figure 1 A fabricated gold/Si gratings x-ray mask.

$$\Lambda = \frac{\lambda_0}{2n_{eff}} = \frac{c_0}{2f\ n_{eff}}, \qquad (1)$$

where $n_{eff}$ is the effective refractive index of the fiber core; $c_0$ velocity of light in vacuum; and $f$ is the frequency of the light.

From Eq. (1), the pitch/period $\Lambda \approx 534.5\,\mathrm{nm}$ for a typical refractive index $n \approx 1.45$ and a vacuum wavelength of $\lambda_0 = 1550\,\mathrm{nm}$.

Also, it is well known that high-energy x-ray photons change the Radiation-Induced Attenuation (RIA) or the imaginary part of the refractive index greatly [10]: As an example, after an x-ray pulse of a few tens of nanoseconds, RIA levels as high as 2000 dB km$^{-1}$ at 1550 nm (to be compared to 0.2 dB km$^{-1}$ before irradiation) have been observed for Corning SMF28 fibers,

The work is partly supported by National Nuclear Security Administration (NNSA) of the U.S. Department of Energy (DOE), under Contract No. DOE-NNSA PRJ1000773, Use of the Center for Nanoscale Materials, and Advanced Photon Source, Office of Science user facilities, was supported by the U.S. DOE, Office of Science, Office of Basic Energy Sciences, under Contract No. DE-AC02-06CH11357.

meaning that 50% of the signal is absorbed in ~1.5 m. What's more, according to the Kramers-Kronig Relation [11], the real part of the refractive index $n_r'$ can be obtained from the imaginary part of the refractive index $n_r''$ as follows,

$$[n_r'(\omega)]^2 = 1 + \frac{2}{\pi} P \int_0^\infty \frac{\omega'[n_r''(\omega')]^2}{(\omega')^2 - (\omega)^2} d\omega', \tag{2}$$

where $P$ denotes the principal integration. From the Kramers-Kronig relation of Eq. (2), it is expected that the real part of the refractive index $n_r'$ will change if the RIA changes, which is closely related to the imaginary part of the efractive index $n_r''$.

## 2. X-ray Writing Gold/Si Gratings Mask

To use x-ray to induce refractive index contrast on the optical fiber, an x-ray intensity mask made of gold (Au) on a Si substrate is required. From Eq. (1), the x-ray intensity grid mask consists of gold/Si grating of a period $\Lambda \approx 534.5\,\text{nm}$, which was done at the Center for Nanoscale Materials (CNM) of Argonne National Laboratory (ANL). To have high x-ray writing contrast of the optical fiber FBGs, high aspect ratio of the gold/Si gratings mask is required, which is very challenging [12]. To accomplish this, the gold/Si gratings mask fabrication involved e-beam lithography, deep reactive ion etching, atomic layer deposition, and Au electroplating. Fig. 1 shows the fabricated gold/Si gratings from which the high aspect ratio (>10) of the gratings can be seen.

## 3. X-ray Writing of the Optical Fiber FBGs

After the nanoscale x-ray intensity mask was fabricated at CNM of ANL, the fiber FBGs x-ray writing work was carried out at Sector 1 of the Advanced Photon Source (APS) at ANL. The setup of the x-ray writing of the optical fiber is shown in Fig. 2. During the x-ray writing of optical fiber FGBs, the x-ray beam (1-100 keV, >$10^{12}$ photons/sec/mm$^2$) is focused to a short optical fiber (Fig. 1a) beneath the mask grating (Fig.1b) which will create the x-ray exposure pattern to the optical fiber and modulate its refractive index to form the FBGs. Finally, a laser optics system was used to monitor the x-ray writing process in real time. It consists of a C-band

The work is partly supported by National Nuclear Security Administration (NNSA) of the U.S. Department of Energy (DOE), under Contract No. DOE-NNSA PRJ1000773, Use of the Center for Nanoscale Materials, and Advanced Photon Source, Office of Science user facilities, was supported by the U.S. DOE, Office of Science, Office of Basic Energy Sciences, under Contract No. DE-AC02-06CH11357.

(1530-1565 nm) laser on one end of the optical fiber and a C-band InGaAs photodetector on the other end of the optical fiber to detect the transmission intensity.

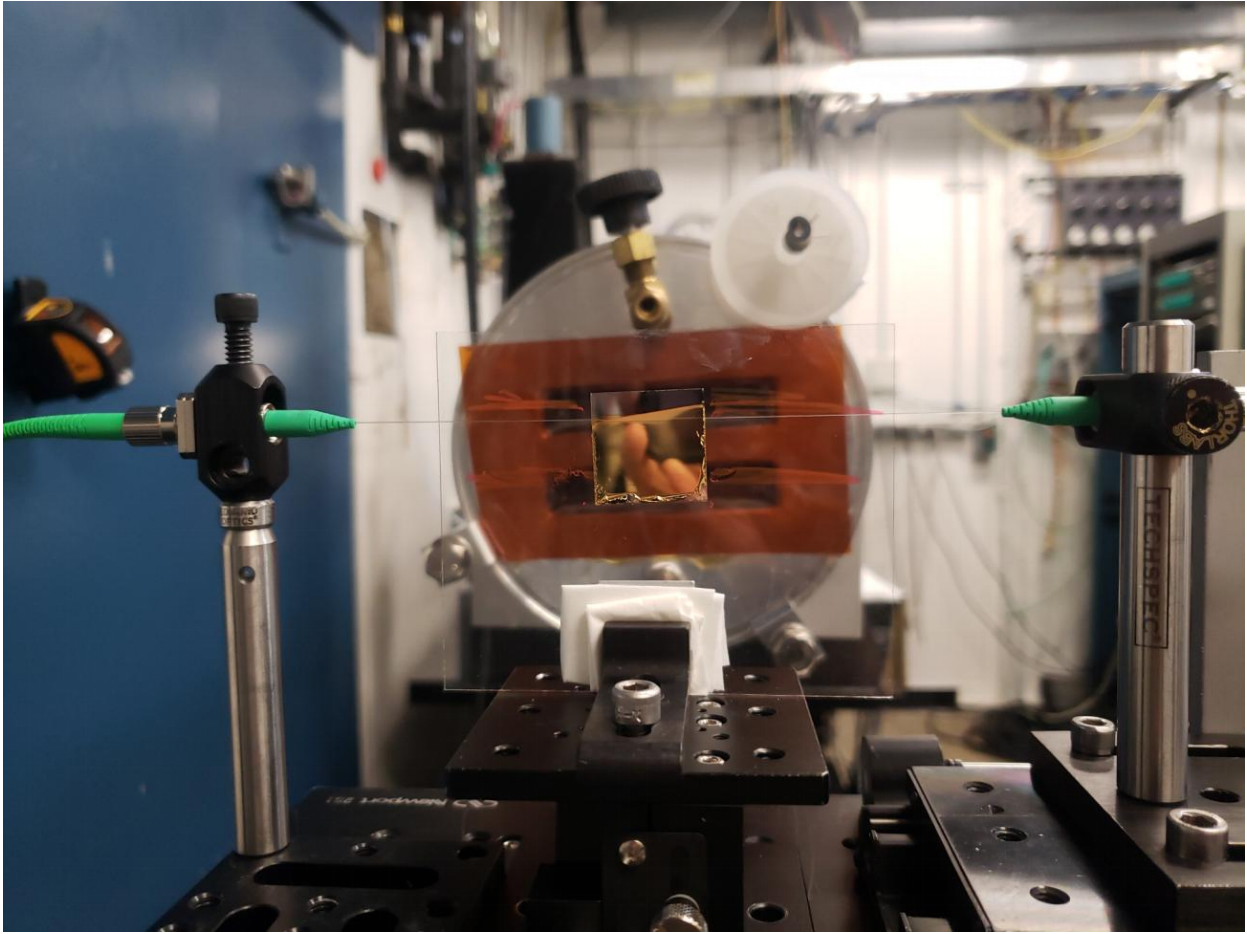

Figure 2 Setup of the x-ray writing of the FBGs at ANL's APS beam line.

## 4. Preliminary Result

Fig. 3 shows the detected transmission signal when the optical fiber is exposed to the synchrotron x-ray source. It can be seen that a significant increase of RIA, i.e., drop of the transmission signal, occurs when the x-ray is on, due to x-ray generated photoelectrons, as well as the chemical bonds change of the optical fiber materials. After the x-ray is turned off, the transmission signal recovers gradually, but cannot recover to the original level before th x-ray is turned on. This is due to the permanent change of the chemical bonds that leads to permanent change of the refractive index the optical fiber. According to the Kramers-Kronig relation of Eq.

The work is partly supported by National Nuclear Security Administration (NNSA) of the U.S. Department of Energy (DOE), under Contract No. DOE-NNSA PRJ1000773, Use of the Center for Nanoscale Materials, and Advanced Photon Source, Office of Science user facilities, was supported by the U.S. DOE, Office of Science, Office of Basic Energy Sciences, under Contract No. DE-AC02-06CH11357.

(2), the real refractive index of the optical fiber also changes with the RIA, indicating that it is promising to write the FBGs using x-ray.

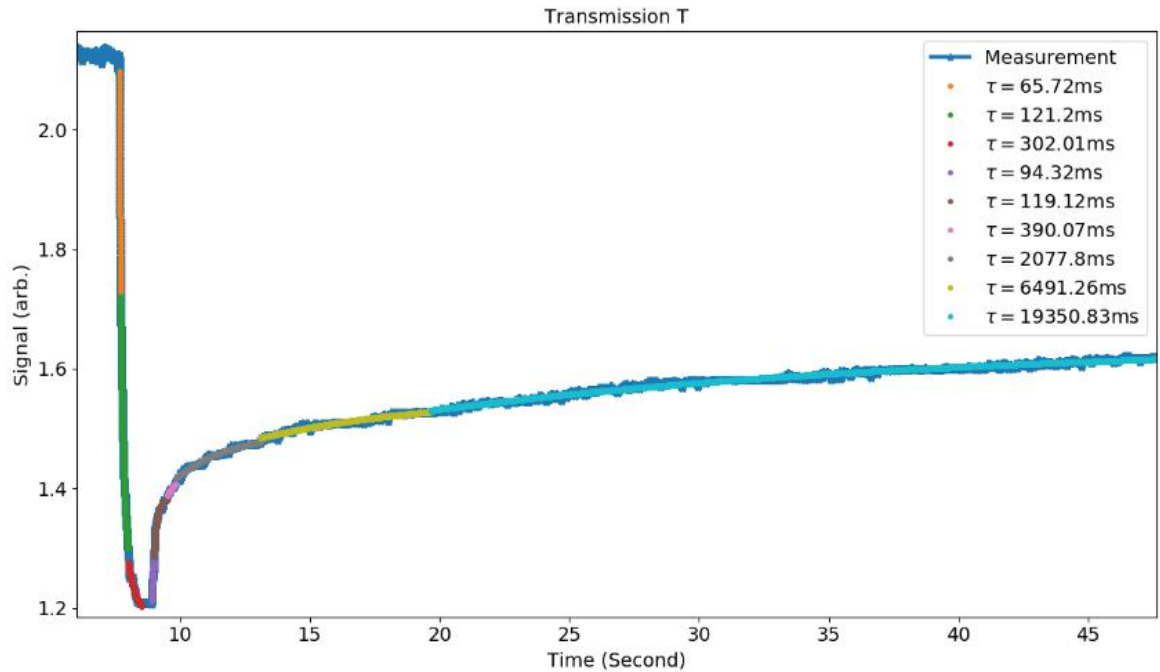

Figure 3 The measured optical fiber transmission and calculated relaxation/recovery times.

## 5. Conclusion

We have investigated a novel method of writing the FBGs using the high-flux x-ray. To form the FBGs on the optical fiber, a nanoscale x-ray intensity mask made of gold/Si gratings is used to generate the x-ray exposure pattern on the optical fiber. Then the x-ray changes chemical bonds of the optical fiber and thus its RIA, which is closely related to the real part of the refractive index according to the Kramers-Kronig relation. Experiment has been carried out at ANL's APS synchrotron x-ray source facility. The preliminary result shows that the x-ray does induce significant RIA change and thus the change of the real part of the refractive index, showing that the fabrication of FBGs with high-flux x-ray is promising.

The work is partly supported by National Nuclear Security Administration (NNSA) of the U.S. Department of Energy (DOE), under Contract No. DOE-NNSA PRJ1000773, Use of the Center for Nanoscale Materials, and Advanced Photon Source, Office of Science user facilities, was supported by the U.S. DOE, Office of Science, Office of Basic Energy Sciences, under Contract No. DE-AC02-06CH11357.

The work is partly supported by National Nuclear Security Administration (NNSA) of the U.S. Department of Energy (DOE), under Contract No. DOE-NNSA PRJ1000773, Use of the Center for Nanoscale Materials, and Advanced Photon Source, Office of Science user facilities, was supported by the U.S. DOE, Office of Science, Office of Basic Energy Sciences, under Contract No. DE-AC02-06CH11357.

The work is partly supported by National Nuclear Security Administration (NNSA) of the U.S. Department of Energy (DOE), under Contract No. DOE-NNSA PRJ1000773, Use of the Center for Nanoscale Materials, and Advanced Photon Source, Office of Science user facilities, was supported by the U.S. DOE, Office of Science, Office of Basic Energy Sciences, under Contract No. DE-AC02-06CH11357.